\documentclass[prodmode,acmcsur]{acmsmall} 

\usepackage{longtable}
\usepackage{rotating}
\usepackage[english]{babel}
\usepackage[utf8]{inputenc}
\usepackage{amsmath}
\usepackage{graphicx}
\usepackage[usenames,dvipsnames]{xcolor}
\usepackage[T1]{fontenc}
\usepackage{url}
\usepackage[numbers]{natbib}
\usepackage[colorinlistoftodos, backgroundcolor=GreenYellow]{todonotes}
\usepackage{url}
\usepackage{booktabs}

\acmVolume{50}
\acmNumber{5}
\acmArticle{69}
\acmYear{2017}
\acmMonth{9}

\doi{10.1145/3108242}

\issn{1234-56789}

\newcommand{\dorien}[1]{\textcolor{blue}{#1}}

\begin{document}

\markboth{D. Herremans et al.}{A Functional Taxonomy of Music Generation Systems}


\title{A Functional Taxonomy of Music Generation Systems}

\author{DORIEN HERREMANS
\affil{Singapore University of Technology and Design \\ \& Queen Mary University of London}
CHING-HUA CHUAN
\affil{University of North Florida}
ELAINE CHEW
\affil{Queen Mary University of London}}

\begin{abstract}



Digital advances have transformed the face of automatic music generation since its beginnings at the dawn of computing. Despite the many breakthroughs, issues such as the musical tasks targeted by different machines and the degree to which they succeed remain open questions. We present a functional taxonomy for music generation systems with reference to existing systems. The taxonomy organizes systems according to the purposes for which they were designed. It also reveals the inter-relatedness amongst the systems. This design-centered approach contrasts with predominant methods-based surveys, and facilitates the identification of grand challenges so as to set the stage for new breakthroughs.

\end{abstract}

\ccsdesc[500]{Applied computing~Sound and music computing}
\ccsdesc[300]{Information systems~Multimedia information systems}
\ccsdesc[300]{Computing methodologies~Artificial intelligence}
\ccsdesc[100]{Computing methodologies~Machine learning}


%
%


\keywords{music generation, taxonomy, functional survey, survey, automatic composition, algorithmic composition}

\acmformat{Dorien Herremans, Ching-Hua Chuan and Elaine Chew, 2016. A Functional Taxonomy of Music Generation Systems.}

\begin{bottomstuff}
This project has received funding from the European Union’s Horizon 2020 research and innovation programme under grant agreement No 658914. 

Author's addresses: D. Herremans, Information Systems Technology and Design Pillar, Singapore University of Technology and Design, Singapore University of Technology \& Design, 8 Somapah Road, 1.502-18, Singapore 487372, for part of the work, D. Herremans was at the School of Electronic Engineering and Computer Science, Queen Mary University of London; E. Chew, School of Electronic Engineering and Computer Science, Queen Mary University of London, Mile End Road, E1 NS4 London, UK; C.-H. Chuan, 	
School of Computing, University of North Florida, 1 UNF Drive, Jacksonville, FL 32224, US.
\end{bottomstuff}

\maketitle

\section{Introduction}
\label{sec:intro}

\noindent
The history of automatic music generation is almost as old as that of computers. That machines can one day generate ``elaborate and scientific pieces of music of any degree of complexity and extent''~\citep{lovelace1843notes} was anticipated by visionaries such as Ada Lovelace since the initial designs for a general purpose computing device were laid down by Charles Babbage. Indeed, music generation or automated composition was a task accomplished by one of the first computers built, the ILLIAC I~\citep{hiller1957musical}. Today, computer-based composition systems are aplenty. 
The recent announcement of Google Magenta\footnote{\url{http://magenta.tensorflow.org/welcome-to-magenta}}, ``a research project to advance the state of the art in machine intelligence for music and art generation,'' underscores the importance and popularity of automatic music generation in artificial intelligence. 

Despite the enthusiasm of researchers, using computers to generate music remains an ill-defined problem. Although several survey papers on automatic music generation~\citep{papadopoulos1999ai, nierhaus2009algorithmic, fernandez2013ai} exist, researchers still debate the kinds of musical tasks that can be performed by machines and the degree to which satisfactory outcomes can be achieved. Outstanding questions include: what compositional tasks are solved and which remain challenges? How is each compositional task modeled and how do they connect to each other? What is the relationship between systems proposed for different compositional tasks? What is the goal defined for each task and how can the objective be quantified? How are the systems evaluated? While individual questions or subsets of these questions might be addressed in specific papers, previous surveys fail to provide a systematic comparison of the state of the art.

This paper aims to answer these questions by proposing a functional taxonomy 
of automatic music generation systems. Focusing on the purpose for which the systems were developed, we examine the manner in which each music composition task was modeled and describe the connection between different tasks within and across systems. We propose a concept map for automatic music generation systems based on the functions of the systems in Section~\ref{sec:architecture}. A brief history of early automatic music generation systems is provided in Section~\ref{sec:early}, 
followed by a discussion on the general approach to evaluating computer generated music (Section~\ref{sec:success}). A detailed survey of systems designed based on each functional aspect is then presented in Section~\ref{sec:aspects}.

\subsection{Function and design concepts in automatic music generation systems}
\label{sec:architecture}

The complexity and types of music generation systems is almost as varied as music itself. It would be a gross simplification to consider and judge all automatic music generation systems in a homogeneous fashion. The easiest way to understand the complexity of these systems and their connections  one to another is to examine the functions for which they were designed.

\begin{figure*}[t]\centering
\includegraphics[width = 0.8\linewidth]{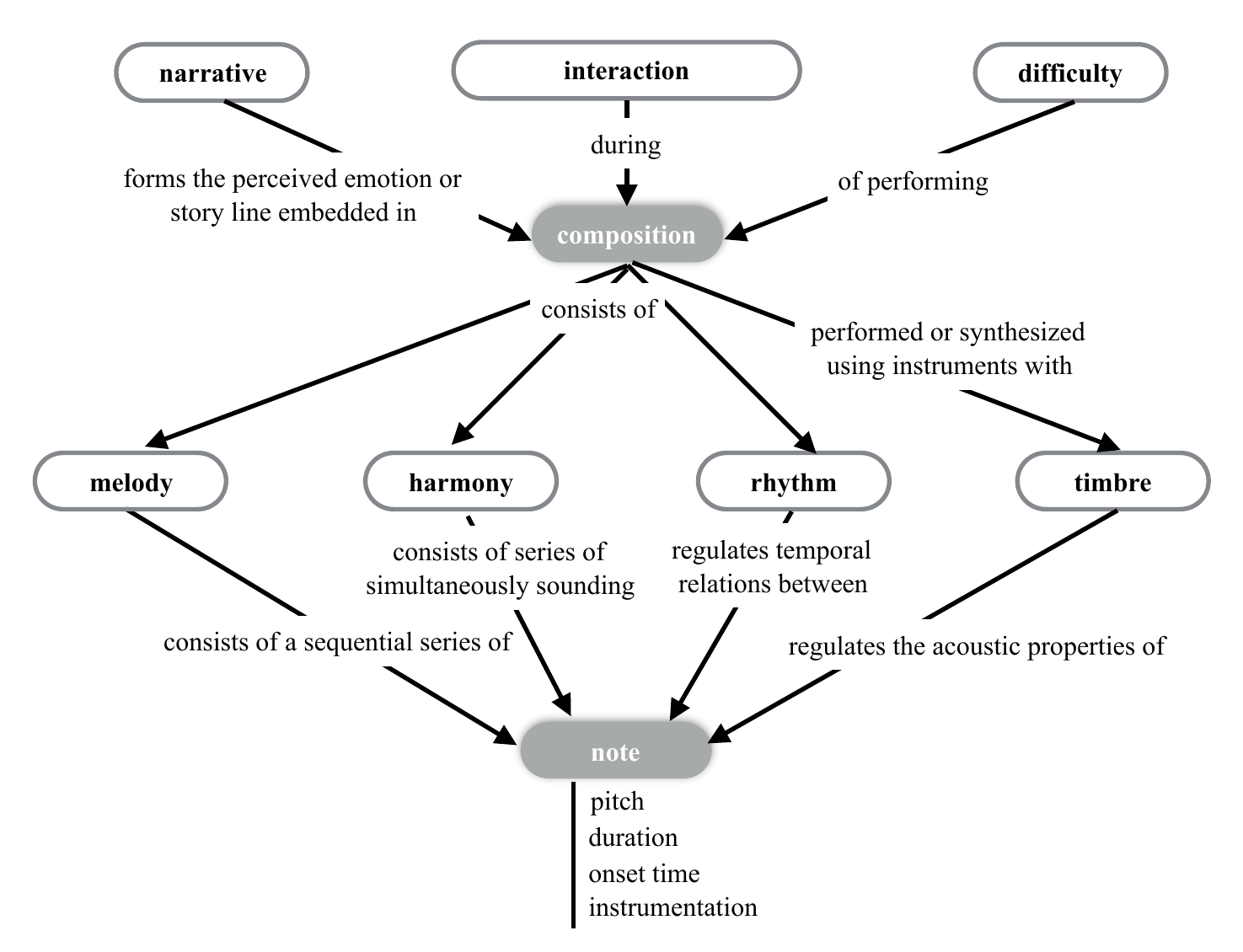}
\caption{Concept map for automatic music generation systems.}
\label{fig:concept}
\end{figure*}

Figure~\ref{fig:concept} illustrates a concept map showing the functional design aspects that form the proposed taxonomy of music generation systems. The map is centered around two basic concepts crucial to music generation systems: the {\em composition} (the higher grey node) and the {\em note} (the lower gray node), which possesses properties such as pitch, duration, onset time, and instrumentation. 

Between the {\em note} and the {\em composition} lie four essential elements of music composition: {\em melody}, {\em harmony}, {\em rhythm}, and {\em timbre}. Systems that focus on any one of the four aspects generate a sequence of notes that fulfills a specific set of goals, which can vary widely amongst the systems. For example, for melody generation, a system could be designed to simply produce a monophonic sequence of notes~\citep{brooks1957experiment}, or be constrained to fit a given accompaniment~\citep{pachet2001finite}. For an automatic harmonization system, the goal could involve  generating three lines of music for a given melody without breaking music theoretic rules (e.g., harmonizing chorales~\citep{ebcioglu1988expert}, or producing substitute chord progressions in jazz~\citep{chemillier2001improvising}. For rhythm generation, a system could focus on producing rhythmic patterns that sound like rock n' roll~\citep{tokui2000music}, or on changing the timing of onsets to make the rendering of the piece sound more human-like~\citep{tidemann2008drum}. 

Timbre is unique in that it is based only on the acoustic characteristic of music. Timbre can be generated either by playing notes on a real instrument or by artificially synthesizing sounds for a note or several notes. In automatic music composition, timbre generation surfaces as a problem in orchestration, which is often modeled as a retrieval problem~\citep{psenicka2003sporch}, or a multi-objective search problem~\citep{carpentier2010solving}.  

The objective of a system, such as matching a target timbre, will directly impact the problem definition and prospective solution techniques, such as multi-objective search or retrieval. 
Also notice that a music generation system can tackle more than one functional aspect---melody, harmony, rhythm, timbre---either by targeting multiple goals at the same time or focusing on one goal with other musical aspects considered constant and provided by the user.

Returning to Figure~\ref{fig:concept}, three high-level concepts are shown above {\em composition}: {\em narrative}, {\em interactive composing}, and {\em difficulty}. Interactive composing refers to an online problem solving approach, which can be real-time or not, to music generation that employs user input. A system can be designed to generate each of the four essential musical elements, or a combination of them, in an interactive manner. For example, a system can listen to a person's playing and learn her or his style in real time, and improvise with the player in the same style ~\citep{pachet2003continuator, assayag2006omax}. Another type of interactive system incorporates a user's feedback in the music generation process,  using it either as critique for reinforcement learning~\citep{franklin2001multi} or as a source of parameters in music generation~\citep{franccois2013mimi4x}.

The {\em narrative} contributes to the emotion, tension, and/or story line perceived by the listener when listening to music~\citep{huron2006sweet}. 
The concept of {\em difficulty} focuses on physical aspects of playing the instrument. Systems with ergonomic goals must consider the playability of certain note combinations on a particular instrument. 

To achieve these goals, the long-term and/or hierarchical structure of the music plays an important role. These high-level goals and the long-term structure have been the focus of recent development in automatic music generation, a trend that will persist into the near future.



As shown in Figure~\ref{fig:concept}, automatic music generation evokes a number of computational problems and demonstrates capabilities that span almost the entire spectrum of artificial intelligence. For example, generating music can be described as a sensorless problem (generating monophonic melody without accompaniment), a partially observable problem (with accompaniment but not the underlying chord progression), or a fully observable problem (accompaniment with labeled chord progression). Different agent types, including model- and knowledge-based ~\citep{chuan2011generating}, goal-based ~\citep{pachet2001finite}, utility-based~\citep{mcvicarautoguitartab}, and statistical learning~\citep{tokui2000music}, have been used for music generation. 

In music generation, states can be defined in terms of discrete (e.g., pitch, interval, duration, chord) as well as continuous (e.g., melodic contour, acoustic brightness and roughness) features. In addition, various techniques, such as stochastic approaches, probabilistic modeling, and combinatorial optimization, have been applied to music generation. In such a rich problem domain, it is thus especially important to understand the intricacies within each subproblem and the manner in which the subproblems are interconnected one with another. 


\subsection{Automating composition: early years}
\label{sec:early}


The idea of composers relinquishing some degree of creative control and automating certain aspects of composition has been around for a long time. A popular early example is Mozart's {\em Musikalisches W\"urfelspiel} (Musical Dice Game), whereby small fragments of music are randomly re-ordered by rolling a dice to create a musical piece. Mozart was not the only one experimenting with this idea. In fact, the first musical dice game, called Der allezeit fertige Menuetten und Polonaisencomponist (The Ever-Ready Minuet and Polonaise Composer) can be traced back to Johann Philipp Kirnberger~\citep{kirnberger1757allezeit}. According to~\citet{hedges1978dice}, at least twenty musical dice games where published between 1757 and 1812, making it possible for musical novices to compose polonaises, minuets, marches, walzes, and more. 

John Cage, Charles Dodge, Iannis Xenakis and other avant-garde composers have continued the ideas of chance-inspired composition. John Cage's {\em Atlas Eclipticalis} was composed by randomly placing translucent paper on a star chart and tracing the stars as notes~\citep{pritchett1994completion}. In the piece called ``Analogique A'', Xenakis uses statistical models (Markov) to determine how musical sections are ordered~\citep{xenakis1992formalized}. The composer David Cope began his ``Experiments in Musical Intelligence'' in 1981 as the result of a composer's block; the aim of his resultant software was to model his own composing style, so that at any given point one could request a next note, next bar, and so on. In later experiments, Cope also modeled styles of other composers~\citep{cope1996experiments}. Some of the music composed using this approach proved to be fairly successful. A more extensive overview of such avant-garde composers is given by~\citet{cope2000new}.

State-of-the-art music generation systems extend these ideas of mimicking styles and pieces, be it in the form of statistical properties of styles or explicitly repeated fragments. Before Section~\ref{sec:aspects} describes in greater depth music generation systems in terms of their functions, the next section focuses on how the goals of music generation systems are defined and evaluated, depending on the technique used for generation.

\subsection{Measuring success}
\label{sec:success}

For automatic music generation systems, unless the end goal is the process rather than the outcome, evaluation of the resulting composition is usually desired, and for some systems an essential step in the composition process. 

The output of music generation systems can be evaluated by human listeners, using music theoretic rules, or using machine-learned models. The choice of evaluation method is primarily influenced by the goal of the music generation system, such as similarity to a corpus or a style (as encapsulated by rules or machine-learned models) versus music that sounds good. All of these goals are interrelated and impact the implementation of the music generation system and the quality of the generated pieces.

While human feedback may arguably be the most sound approach for evaluating post-hoc if the generated pieces sound good~\citep{pearce2001towards, agres2015creativity}, requiring people to rate the output at each step of the process can take an excessive amount of time. This is often referred to as the human fitness bottleneck~\citep{biles2001autonomous}. A second issue with human evaluation is fatigue. Continuous listening and evaluating can cause significant psychological strain for the listener~\citep{tokui2000music}. So while it can be useful, and arguably essential, to let human listeners test the final outcome of a system, human ratings aren't practically possible to guide or steer \emph{during} the generation process. 

If the goal of the automatic composition process is to create music similar to a given style or body of work by a particular composer, one could look to music theory for well-known rules such as those for music in the style of a composer, say Palestrina. These could be incorporated into an expert system or serve as a fitness function, say of a genetic algorithm. The downside to this approach is that existing rule sets are limited to a few narrowly-defined styles that have been comprehensively analyzed by music theorists or systematically outlined by the composer, which constrains its robustness and wider applicability, or are so generic as to result in music lacking definable characteristics.

The third approach, using machine-learned models, seems to offer a solution to the aforementioned problems. By learning the style of either a corpus of music or a particular piece, music can be generated with characteristics following those in the training pieces. The characteristics may include distributions of absolute or relative pitch sets, durations, intervals, and contours. A large collection of features is suggested by~\citet{towsey2001towards} and~\citet{conklin95}. 
Markov chains form a class of machine-learned models; they capture the statistical occurrence of features in a particular piece or corpus. Sampling from Markov models results in pieces with similar statistical distributions of the desired musical features. Other machine-learning approaches include neural networks and, more recently, deep-learning methods, which attempt to capture more complex relationships in a music piece. 

A concept that directly relates to the task of evaluating the generated music, regardless of which of the  above three methods are used, is \emph{similarity}. In the first case, human listeners have formed their frame of reference through previous listening experiences \citep{peretz1998exposure, krumhansl2001cognitive} and will judge generated pieces based on their similarity to pieces with which they are familiar.
Secondly, a piece generated with music theoretic rules will possess attributes characteristic of those in the target style. Finally, pieces generated by machine-learned models will have features distributed in ways similar to the original corpus.

Since similarity is central to metrics of success in music generation systems, an important challenge then becomes one of finding the right balance between similarity and novelty or creativity. In the words of \citet{hiller1989lejaren}: ``It is interesting to speculate how much must be changed to create a new work.'' For example, music based on fragments of an already existing composition, as in the case with high-order Markov models, run the risk of crossing the fine line between stylistic similarity and plagiarism~\citep{papadopoulos2014avoiding}. 
Evaluating the creativity, which is sometimes equated to novelty, of the generated music is a complex topic treated in greater length in~\citet{agres2015creativity}. 

In order to facilitate the comparison of results from different music generation systems, the authors have set up an online computer generated music repository\footnote{\url{http://dorienherremans.com/cogemur}}. This repository allows researchers to upload both audio files and sheet music generated by their systems. This will facilitate dissemination of results and promote research transparency so as to better assess the impact of different systems. Access to concrete examples through the website will allow visitors to better understand the behavior of the music generation systems that created them.


In the remainder of this paper, we will discuss each of the functional areas on which music generation systems can focus. Rather than aiming to provide an exhaustive list of music generation systems, we choose to focus on research that presented novel ideas which were later adopted and extended by other researchers. This function and design-based perspective stands in contrast to existing survey papers, which typically categorize generation systems according to the techniques that they employ, such as Markov models, genetic algorithms, rule-based systems, and neural networks---see~\citep{papadopoulos1999ai, nierhaus2009algorithmic, fernandez2013ai}. 
By offering a new taxonomy inspired by the function and design foci of the systems, we aim to provide deeper insights into the problems that existing systems tackle and the current challenges in the field, thereby inspiring future work that pushes the boundaries of the state-of-the-art.

\section{A functional index of music generation systems}
\label{sec:aspects}

This section explores functional aspects addressed in different music generation systems which form the taxonomy proposed in this paper; example systems are given for each aspect. The functional aspects discussed, in order of appearance, are {\em melody}, {\em harmony}, {\em rhythm}, {\em timbre}, {\em interaction}, {\em narrative}, and {\em difficulty}. We also touch upon {\em long-term structure} in relation to some of these categories. 

It is worth pointing out that the aspects, while separate in their own right, can often be conflated; for example, rhythm is inherent in most melodies. Therefore, a system mentioned in the context of one aspect may also touch upon other functional aspects.

In Table~\ref{tab:overview} an overview is given of the different techniques used within these functional aspects. Systems are classified by their main technique and listed with their most prominent aspect. Typically, music generation systems can belong to more than one category. In this paper (and therefore also in Table~\ref{tab:overview}), the most important contribution of the systems is emphasized and only the systems with a clear contribution are listed. In the next subsections, the individual functional aspects will be discussed in greater detail. 

\small
\renewcommand{\arraystretch}{1.4} 

\begin{longtable}{p{50pt} p{330pt}}  
\caption{Functional overview of selected music generation systems by their main technique.}
\label{tab:overview}
\endfirsthead
\endhead
\toprule
\multicolumn{2}{c}{\textbf{Markov models}} \\
\midrule
Melody  & \cite{pinkerton1956information, brooks1957experiment, moorer1972music, conklin95, pachet2001finite, davismoon2010combining, pearce2010unsupervised,  gillick2010machine, mcvicarautoguitartab, papadopoulos2014avoiding} \\

Harmony &
\cite{hiller1957musical, xenakis1992formalized, farbood2001analysis, allan2004harmonising, lee2004ring, Yi2007Automatic, simon2008mysong, eigenfeldt2009realtime, de2010neural, chuan2011generating, bigo2015viewpoint} \\

Rhythm & \cite{tidemann2008drum,  marchini2010unsupervised, hawryshkewich2011beatback} \\ 

Interaction & \cite{thom2000bob} \\

Narrative & \cite{prechtl2014algorithmic, prechtl2014methodological} \\

Difficulty & \cite{mcvicarautoguitartab} \\

\midrule
\multicolumn{2}{c}{\textbf{Factor oracles}} \\
\midrule

Interaction & 
\cite{assayag2006omax, weinberg2006toward, franccois2007visual, assayag2010interaction, dubnov2012music, franccois2013mimi4x, nika2015guided} \\

Rhythm & \cite{weinberg2006toward} \\

\midrule
\multicolumn{2}{c}{\textbf{Incremental parsing}} \\
\midrule

Interaction & \cite{pachet2003continuator}\\

\midrule
\multicolumn{2}{c}{\textbf{Reinforcement learning}} \\
\midrule

Interaction & \cite{franklin2001multi} \\

\midrule
\multicolumn{2}{c}{\textbf{Rule/Constraint satisfaction/Grammar-based}} \\
\midrule

Melody & 
\cite{keller2007grammatical, gillick2010machine, herremans2012composing} \\

Harmony &
\cite{hiller1957musical,  steedman1984generative, ebcioglu1988expert, cope1996experiments, assayag1999computer, cope2004musical, huang2005palestrina, anders2007composing, anders2009computational, aguilera2010automated, herremans2012composing, herremans2013composing, tanaka2016babbitt} \\

Narrative &
\cite{rutherford2002experiment} \\

Difficulty &
\cite{lin2006intelligent} \\

Interaction &
\cite{lewis2000too, chemillier2001improvising, morales2001sicib, marsden2004novagen} \\

Narrative & 
\cite{casella2001magenta, farbood2007composing, brown2012mezzo, nakamura1994automatic} \\

\midrule
\multicolumn{2}{c}{\textbf{Neural networks/Restricted Boltzmann machines/ LSTM}} \\
\midrule

Harmony & 
\cite{music1991creation, hild1992harmonet,  eck2002first,  ICML2012BoulangerLewandowski_590,herremans2017word2vec} \\

Melody & 
\cite{todd1989connectionist, duff1989backpropagation, mozer1991connectionist, music1991creation, toiviainen1995modeling, eck2002first,franklin2006recurrent, agres2009sparsity, ICML2012BoulangerLewandowski_590} \\

Interaction &
\cite{franklin2001multi} \\

Narrative & 
\cite{browne2009global} \\


\midrule
\multicolumn{2}{c}{\textbf{Evolutionary/Population-based optimization algorithms}} \\
\midrule

Melody & \cite{horner1991genetic, towsey2001towards, waschka2007composing, herremans2012composing} \\

Harmony & \cite{mcintyre1994bach,  polito1997musica, phon1999four, geis2007ant, waschka2007composing, herremans2012composing} \\

Rhythm & \cite{tokui2000music, pearce2001towards, ariza2002prokaryotic} \\

Interaction &
\cite{biles1998interactive, biles2001autonomous} \\

Difficulty & 
\cite{tuohy2005genetic, de2012differential} \\

Timbre &
\cite{carpentier2010solving} \\

\midrule
\multicolumn{2}{c}{\textbf{Local search-based optimization}} \\
\midrule

Melody & 
\cite{herremans2012composing} \\

Harmony & 
\cite{herremans2012composing, herremans2013} \\

Narrative & 
\cite{browne2009global, herremans16orbel} \\

Timbre & 
\cite{carpentier2010solving} \\

\midrule
\multicolumn{2}{c}{\textbf{Integer Programming}} \\
\midrule

Melody & \cite{cunha2016} \\

\midrule
\multicolumn{2}{c}{\textbf{Other optimization methods}} \\
\midrule

Melody & \cite{davismoon2010combining} \\

Harmony & \cite{tsang1991harmonizing, farbood2001analysis, bemman2016generating} \\

Timbre & \cite{hummel2005simulation, collins2012automatic} \\

Difficulty & \cite{radi2004path} \\

\bottomrule

\end{longtable}

\dorien{
}


\normalsize

\subsection{Melody}
\label{sec:melody}

Melody constitutes one of the first aspects of music subject to automatic generation. This section explores the range of automatic systems for generating melody. The generation of simple melodies is studied first, followed by the transformation of existing ones, then the more constrained problem of generating melodies that fit an accompaniment or chord sequence.

\subsubsection{Melodic generation}
\label{sec:melodygen}

When considering the problem of generating music, the simplest form of the exercise that comes to mind is the composition of monophonic melodies without accompaniment.

\paragraph{Problem description}
In most melody generation systems, the objective is to compose melodies with characteristics similar to a chosen style---such as Western tonal music or free jazz---or corpus---such as music for the Ethiopian lyre the bagana~\citep{herremans2015generating}, a selection of nursery rhymes~\citep{pinkerton1956information}, or hymn tunes~\citep{brooks1957experiment}.

These systems depend on a function to evaluate the fitness of output sequences or to prune candidates. Such a fitness function, as discussed in Section~\ref{sec:success} is often based on similarity to a given corpus, style, or piece. The music is often reduced to extracted features; these features can then be compared to that of the exemplar piece or corpus, a model, or existing music theoretic rules. Example features include absolute or relative pitch~\citep{Conklin03}, intervals~\citep{herremans2013}, durations~\citep{Conklin03}, and contours~\citep{alpern1995techniques}. Not all studies provide details of the extracted features, which makes it difficult to compare the objectives and results.



\paragraph{Early work}
Building on the ideas of the aforementioned avant garde composers, some early work on melody generation uses stochastic models. These models capture the statistical occurrence of features in a particular song or corpus to generate music having selected feature distributions similar to the target song or corpus.

The first attempts at generating melodies with computers date back to 1956, when Pinkerton built a first order Markov model, the ``Banal Tune-Maker'', based on a corpus of 39 simple nursery rhymes. Using a random walk process, he was able to generate new melodies that ``sound like nursery rhymes''. The following year, \citet{brooks1957experiment} built Markov models from order one up to eight based on a dataset of 37 hymns. When using a random walk process, they noted that melodies generated by higher order models tend to be more repetitive and those generated by lower order models had more randomness. 

The trade-off between composing pieces similar to existing work and novel, creative input is a delicate one. Although Stravinsky is famously quoted as having said, ``good composers borrow and great composers steal''~\citep{raines2015composition}, machines still lack the ability to distinguish between artful stealing and outright plagiarism. Concepts of irony and humor can also be difficult to quantify. In order to avoid plagiarism and create new and original compositions, an automatic music generation system needs to find the balance between generating pieces similar to a given style, yet not too similar to individual pieces.

\citet{papadopoulos2014avoiding} examined problems of plagiarism arising from higher order Markov chains. Their resulting system learns a high order model, but introduces MaxOrder, the maximum allowable subsequence order in a generated sequence, to curb excessive repeats of material from the source music piece. The sequences are generated using finite-domain constraint satisfaction. The idea of adding control constraints when generating music using Markov models was further explored by~\citet{pachet2001finite}. Examples of applications of such control constraints include requirements that a sequence be globally ascending or follows an arbitrary pitch contour. 
Although there have been some tests of using control constraints with monophonic melodies, the research of~\citet{pachet2001finite} focuses on the even more constrained problem of generating jazz solos over accompaniment, a topic that is explored in greater detail in  Section~\ref{sec:jazz}.


\paragraph{Structure and patterns}
Composing a monophonic melody may seem like a simple task compared to the scoring of a full symphony. Nevertheless, melodies are more than just movements between notes, they normally possess long term structure. This structure may result from the presence of motives, patterns, and variations of the patterns. Generating music from a Markov model with a random walk or Gibbs sampling typically does not enforce patterns that lead to long term structure. In recent years, some research has shown the effectiveness of using techniques such as optimization and deep learning to enforce long-term structure. 


\citet{davismoon2010combining} were some of the first researchers to frame music generation as a combinatorial optimization problem with a Markov model integrated in its objective function. In order to evaluate the music generated, their system builds a (second) Markov model based on the generated music so as to enable to system to minimize a Euclidean distance between the original model and the new model. They used simulated annealing, a metaheuristic inspired by a metallurgic technique used to cool a crystalline solid~\citep{kirkpatrick1983optimization}, to solve this distance-minimization problem. This allowed them to pose some extra constraints to control pitch drift and solve end-point problems. 

\citet{pearce2010unsupervised}'s IDyOM system uses a combination of long- and short-term Markov models. A dataset of modern Western tonal-style music was used to train a long-term model, combined with a short-term model trained incrementally on the piece being generated. The short-term model captures the changes in melodic expectation as it relates to the growing knowledge of the current fragment's structure. Local repeated structures are more likely to recur; this model will therefore recognize and stimulate repeated structures within a piece. The result is an increase in the similarity of the piece with itself, which can be considered a precursor to form.

A recent study by~\citet{roig2014automatic} generates melodies by concatenating rhythmic and melodic patterns sampled from a database. Selection is done based on rules combined with a probabilistic method. This approach allows the system to generate melodies with larger-scale structure such as repeated patterns, which causes the piece to have moments of self-similarity. \citet{cunha2016} adopt a similar approach, using integer programming with structural constraints to generate guitar solos from short existing licks. The objective function consists of a combination of rules.  
\citet{bemman2016generating} mathematically formalized a problem posed by composer Milton Babbitt. Babbit is famous for composing twelve-tone serial music and formulated the ``all-partition array''-problem, which consists of finding a rectangular area of pitch class integers that can be partitioned into regions whereby each region represents a distinct integer partition of 12. There are only very few solutions to this computationally hard composition problem with a very structured nature, one of which was found by~\citet{tanaka2016babbitt} through constraint programming.

\citet{herremans2015generating} investigates the integration of Markov models in an optimization algorithm, exploring multiple ways in which a Markov model can be used to construct an objective function that forces the music to have the same statistical distribution of features as a corpus or piece. This optimization problem is solved using a variable neighborhood search (VNS). The main advantage of this approach is that it allows for the inclusion of any type of constraint. In their paper, the generated piece is constrained to an AABCA structure. The approach was implemented and evaluated by generating music for the bagana, an Ethiopian lyre. Since this system uses the semiotic pattern from a template piece, the newly generated pieces can be considered as having structure like the template. 

The MorpheuS system~\citep{herremans16orbel} expands on the VNS method, adding constraints on recurring (transposed) patterns and adherence to a given tension profile. Repeated patterns are detected using the compression algorithm COSIATEC~\citep{meredith2013cosiatec}. COSIATEC finds the locations where melodic fragments are repeated in a template piece, thus supplying higher-level information about repetitions and structural organization. Tonal tension is quantified using measures~\citep{herremans16tension} based on the spiral array~\citep{chew2013tonality}.

In recent years, more complex deep learning models such as recursive neural networks have gained in popularity. The trend is due in part to the fact that such models can learn complex relationships between notes given a large-enough corpus. Some of these models also allow for the generation of music with repeating patterns and notions of structure. The next paragraphs examine research on neural network-based melody generation.

\paragraph{Deep learning and structure}
The first computational model based on artificial neural networks (ANNs) was created by~\citet{mcculloch1943logical}. Starting in the eighties, more sophisticated models have emerged that aim to more accurately capture complex properties of music. The first neural network for music generation was developed by~\citet{todd1989connectionist}, who designed a three-layered recurrent artificial neural network, whose output (one single pitch at a time) forms a melody line. Building on this approach,~\citet{duff1989backpropagation} created another ANN using relative pitches instead of absolute pitches to compose music in J.S. Bach's style. Recurrent neural networks are a family of neural networks built for representing sequences~\citep{rumelhart1988learning}. They have cyclic connections between nodes that create a memory structure. 

\citet{mozer1991connectionist} implemented a recurrent connectionist network (called CONCERT), that was used in an experiment to generate music that sounds like J.S. Bach's minuets and marches. Novel in this approach was the representation of pitches in a psychologically-grounded multidimensional space. This representation enabled the system to capture a notion of similarity between pitches. Although CONCERT is able to learn some structure, such as that of diatonic scales, its output lacks long-term coherence such as that produced by repetition and the statement of the theme at the beginning and its return near the end. While the internal memory of recursive neural networks~\citep{rumelhart1985learning} can, in principle, deal with the entire sequence history. It remains a challenge, however, to efficiently train long term dependencies~\citep{bengio1994learning}. 
x
In the same year,~\citet{music1991creation} designed another ANN framework with a slightly different approach. Instead of training the ANN on a corpus, he mapped a collection of patterns---drawn from music ranging from random to very good---to a musicality score. To create new pieces, the mapping was inverted and the musicality score of random patterns was maximized with a gradient-descent algorithm to reshape the patterns. Due to the high computational cost, the system was only tested on simple and short compositions. \citet{agres2009sparsity} built a recurrent neural network that learned the tonal structure of melodies, and examined the impact of the number of epochs of training on the quality of newly generated melodies. They showed that better-liked melodies were the result of models that had more sparse internal representations. Conceptually, this sort of sparse representation may reflect the way in which the human cortex encodes musical structure. 

Since these initial studies, deep learning networks have increased in popularity. \citet{franklin2006recurrent} developed a Long Short-Term Recurrent Neural Network (LSTM) that generates solos over a reharmonization of chords. She suggests that hierarchical LSTM networks might be able to learn sub-phrase structures in future work. LSTM was developed in 1997 by~\citep{hochreiter1997long}. It is a recurrent neural network architecture that introduces a memory structure in its nodes. More recently,~\citet{ICML2012BoulangerLewandowski_590} used a piano roll representation to create Recurrent Temporal Restrictive Boltzmann Machine (RT-RBM)-based models for polyphonic pieces. An RBM, originally called Harmonium by the original developer~\citep{Smolensky1986parallel}, is a type of neural network that can learn a probability distribution over its inputs. While the model of~\citet{ICML2012BoulangerLewandowski_590} is intended mainly to improve the accuracy of transcription, it can equally be used for generating  music. The RBM-based model learns basic harmony and melody, and local temporal coherence. Long-term structure and musical meter are not captured by this model. 

The capability for RBM's to recognize long-term structures such as motives and phrases is acknowledged in a recent paper by~\citet{lattner2015probabilistic}, in which an RBM is used to segment musical pieces. The model reaches an accuracy rate that competes with current state-of-the-art segmentation models. Recent work by \citet{herremans2017word2vec} takes a different approach inspired by linguistics. They use neural networks to evaluate the ability of semantic vector space models (word2vec) to capture musical context and semantic similarity. The results are promising and show that musical knowledge such as tonality can be modeled by solely looking at the context of a musical segment. 


\subsubsection{Transformation}

\citet{horner1991genetic}, pioneers in applying genetic algorithms (GAs) to music composition, tackle the problem of thematic bridging, the transformation of an initial musical pattern to a final one over a specified duration. A genetic algorithms is a type of metaheuristic that became popular in the 70s through the work of~\citep{holland1992adaptation}. It typically maintain a set (called population) of solutions and combine solutions from this set to form new ones. In the work of~\citet{horner1991genetic}, based on a set of operators, an initial melodic pattern is transformed to resemble the final pattern using a GA. The final result consists of a concatenation of all patterns encountered during this process. 

\citet{ralley1995genetic} uses the same technique (GA) for melodic development, a process in which key characteristics of a given melody are transformed to generate new material. The results are mixed as no interesting transformed output were found. According to~\citet{ralley1995genetic}, the problem lies in the high subjectivity of the desired outcome. 

GenDash, a compositional tool developed by composer Rodney Waschka~II \citep{waschka2007composing}, is not a fully automated composition system but works in tandem with a human composer. The genetic algorithm does not have any type of fitness function (human or other); it simply evolves measures of music at random. In this process, each measure is treated as a different population for evolution. Using GenDash, Waschka composed the opera {\em Sappho's Breath} by using a population that consists of twenty-six measures from typical Greek and Medieval songs~\citep{dostal2013evolutionary}.

Recently, Sony Computer Science Labs' Flow Composer has been used to reorchestrate Ode to Joy, the European Anthem, in seven different styles, including Bach chorales and Penny Lane by The Beatles~\citep{pachet2016joyful}. The reorchestrations are based on max-entropy models, which are often used in fields such as physics and biology to model probability distributions with observed pairwise correlations ~\citep{lezon2006using}.


\subsubsection{Chord constraints}
\label{sec:jazz}

A melody is most often paired either with counterpoint or with chords that harmonize the melody. While there exists much work on generating chords given a melody (see Section~\ref{sec:harm}), some studies focus on generating a melody that fit a chord sequence. 

\citet{moorer1972music}, for instance, first generates a chord sequence, then a melodic line against the sequence. The melody notes are restricted to only those in the corresponding chord at any given point in time. At each point, a decision is made, based on a second-order Markov model, to invert melodic fragments based on the chord, or to copy the previous one. The resulting short melodies have a strangely alien sound, which the author attributes to the fact that the ``plan'' or approach is not one that humans use, and the system does not discriminate against unfamiliar sequences.


The generation of jazz solos over an accompaniment is a popular problem~\citep{pachet2001finite, toiviainen1995modeling, keller2007grammatical}. 
The improvisation system (Impro-Visor) designed by~\citet{keller2007grammatical} uses probabilistic grammars to generate jazz solos. The model successfully learns the style of a composer, as reflected in an experiment described by~\citet{gillick2010machine}, where human listeners correctly matched 95\% of solos composed by Impro-Visor in the style of the famous performer Clifford Brown to the original solo. The accuracy was 90\% for Miles Davis, and slightly less, 85\% for Freddie Hubbard. They state that ``The combination of contours and note categories seems to balance similarity and novelty sufficiently well to be characterized as jazz''. The system does not capture long-term structure, which the authors suggest might be solved by using the structure of an existing solo as a template.

\citet{eck2002first} tackle a similar problem, the generation of a blues melody following the generation of a chord sequence. They use a Long Short Term Memory RNN, which the authors claim handles long-term structure well. However, the paper does not provide examples of output for the evaluation of the long-term structure. 

In the next section, we review music generation systems that focus on harmony.

\subsection{Harmony}
\label{sec:harmony}
Besides melody, harmony is another popular aspect for automatic music generation. This section describes automatic systems for harmony generation, focusing on the manner in which harmonic elements such as chords and cadences are computationally modeled and produced in accordance to a specific style. 

In the generation of harmonic sequences, the quality of the output depends primarily on similarity to a target style. For example, in chorale harmonization, this similarity is defined explicitly by adherence to voice-leading rules. In popular music, where chord progressions function primarily as accompaniment to a melody, the desired harmonic progression is achieved mostly by producing patterns similar to existing examples having the same context. The context is defined by the vertical relation between melody and harmony (i.e., notes sounding at the same time) as well as horizontal patterns of chord transitions (i.e., the relationship of notes over time).

In addition to direct comparisons of harmonic similarity, the output of a chord generation system can also be evaluated under other criteria such as similarity to a genre or to the music of a particular artist. 

The system must generate sequences recognizably in the target genre or belonging to a particular corpus, yet not ``substantially similar'' to it~\citep{liebesman2007using} so as to avoid accusations of plagiarism. It is only a short step from similarity and plagiarism to copyright infringement. On copyright protection of ubiquitous patterns such as harmonic sequences, \citet{gherman2008harmony} argues that: ``When determining whether two musical works are substantially similar $\ldots$ the simple, basic harmony or variation should not be protectable as it is functional $\ldots$. The harmony that goes beyond the triviality of primary tonal level and blocked chords is and should be protectable under copyright law.''

The next sections discuss the task of counterpoint generation, followed by harmonization of chorales, general harmonization, and the generating of chord sequences.

\subsubsection{Counterpoint}

Counterpoint is a specific type of polyphony. It is defined by a strict set of rules that handle the intricacies that occur when writing music that has multiple independent (yet harmonizing) voices~\citep{siddharthan1999music}.

In Gradus Ad Parnassum, a pedagogical volume written in 1725, Johann Fux documented a comprehensive set of rules for composing counterpoint music~\citep{fux1971study}, which forms the basis of counterpoint texts up to the present day. Counterpoint, as defined by Fux, consists of different ``species'', or levels of increasing complexity, which  include more rhythmic possibilities~\citep{norden1969fundamental}. 

\paragraph{Problem description}
The process of generating counterpoint typically begins with a given melody called the \emph{cantus firmus} (``fixed song''). The task is then to compose one or more melody lines against it. As the rules of counterpoint are strictly defined, it is relatively easy to use rules to generate or evaluate if the generated sequence sounds similar to the style of the original counterpoint music. The Palestrina-Pal system developed by~\citet{huang2005palestrina} offers an interactive interface to visualize violations of these harmonic, rhythmic and melodic rules.

Automatic counterpoint composition systems typically handle two to four voices. The systems for generating four-part counterpoint are grouped together with four-part chorale harmonization in the next section because they follow similar rules. The systems and approaches described below handle fewer than four voices.


\paragraph{Approaches}

Three main approaches exist for emulating counterpoint style: the first uses known rules to generate counterpoint; the second uses the rules in an evaluation function of an optimization algorithm; and, the last uses machine learning to capture the style.

In the first category, \citet{hiller1957musical} uses rules for counterpoint to generate the first and second movements of the Illiac Suite. David Cope composes first species counterpoint given a cantus firmus in his system ``Gradus.'' Gradus analyses a set of first species counterpoint examples and learns the best settings for 6 general counterpoint goals or rules. These goals are used to sequentially generate the piece, using a rule-based approach~\citep{cope2004musical}. 

Another system, developed by~\citet{aguilera2010automated} uses logic based on probability rules to generate counterpoint parts in C major, over a fixed cantus firmus. In the generation process, the system evaluates only the harmony characteristics of the counterpoint, but not the melodic aspects. The original theory of Johann Fux contains rules that focus both melodic and harmonic interaction~\citep{fux1971study}. 

The second approach, using counterpoint rules as tools for evaluation, is employed in the system called GPmuse, a GA developed by~\citet{polito1997musica}. GPmuse composes fifth species (mixed rhythm) counterpoint starting from a given cantus firmus. It extracts rules based on the homework problems formulated by Fux and uses the rules to define the fitness functions for the GA. The music generated by GPmuse sounds similar to the original style of counterpoint music. A problem with the system is that some ``obvious'' rules were not defined by Fux, such as the need for the performer (singer) to breathe. Since these rules were not explicitly programmed in GPmusic, one example output contained very long phrases which solely contained eight notes without any rests. 

Strasheela is a generic constraint programming system for composing music. \citet{anders2007composing} uses the Strasheela system to compose first species counterpoint based on six rules from music theory. Other constraint programming languages, such as PWConstraints developed at IRCAM can be used to generate counterpoint, provided the user inputs the correct rules~\citep{assayag1999computer}.

\citet{herremans2012composing} uses a more extensive set of eighteen melodic and fifteen harmonic rules based on Johann Fux's theory to generate a cantus firmus and first species counterpoint. The authors implement the rules in an objective function and optimize (increase) the adherence to these rules using a variable neighborhood search algorithm (VNS). VNS is a combinatorial optimization algorithm based on local search proposed by~\citet{mladenovic1997variable}. \citet{herremans2012composing}'s system was also implemented as a mobile app~\citep{antor13}, and later extended by adding additional rules based on Fux to generate fifth species counterpoint~\citep{herremans2013}.



A final approach to the counterpoint generation problem can be seen in the application of a machine-learning method to Palestrina-style counterpoint. \citet{farbood2001analysis} implemented a Hidden Markov Model to capture the different rules of such counterpoint; they found the resulting music to be ``musical and comparable to those created by a knowledgeable musician.'' Hidden Markov Models, first described by~\citet{baum1966statistical}, are used to model systems that are Markov processes with unobserved (hidden) states, and have since become known for their application in temporal pattern recognition~\citep{yamato1992recognizing}. 

\subsubsection{Harmonizing chorales}

The harmonizing of chorales is one of the most popular music generation tasks pertaining to harmony.  
Chorale harmonization produces highly structured music that has been widely studied in music theory, and a rich body of theoretical knowledge offers clear directions and guidelines for composing in this style.

\paragraph{Problem definition}

The problem of chorale harmonization has been formulated computationally in a variety of different ways. The most common form is to generate three  voices designed to harmonize a given melody, usually the soprano voice~\citep{allan2004harmonising, ebcioglu1988expert, geis2007ant, hild1992harmonet, phon1999four, tsang1991harmonizing, Yi2007Automatic}. 

In contrast, the Bach-in-a-Box system proposed by~\citet{mcintyre1994bach} aims to harmonize a user-created melody, which can form one of any four possible voices. Given a monophonic sequence, the system must generate, using GA, three other notes to form a chord with each melody note while ensuring that the given melodic notes are not mutated 
in the process. The quality of a generated four-part sequence is then measured via fitness functions related to the construction of the chord, the pitch range and motion, the beginnings and endings of chords, smoothness of the chord progressions and chord resolution.


Some systems simplify the process by assuming that all melody notes are chord tones and chords exist at every melody note, i.e. that the polyphony is homophonic~\citep{phon1999four, mcintyre1994bach, Yi2007Automatic}. In many systems, non-chord tones such as passing notes are added as an after-thought following the establishing of the chord progression; others incorporate explicit considerations of non-chord tones in the generation process~\citep{allan2004harmonising, hild1992harmonet}.

\paragraph{Solution approach}

As described in~\citep{hild1992harmonet}, the results of chorale harmonization can be expressed in multiple ways, including: as a harmonic skeleton, a chord skeleton, or as four full parts complete with passing tones. A harmonic skeleton describes the chord progression as a sequence of symbols---such as roman numerals---that represent the functional role of each chord in the progression; the rhythm is implied or considered as given. A chord skeleton shows the constituent notes of each chord without passing tones. Most chorale harmonization systems aim to generate chord skeletons; few cover all three kinds of abstractions.

Some systems generate only a harmonic skeleton. For example, \citet{de2010neural}'s system produces functional harmonizations represented as roman numerals with indications of whether the chord is in the root position or some inversion. The system proposed by \citet{anders2009computational} generates the harmonic backbone without requiring melodic input.

It is worth noting that the terminology for the abstractions are not used consistently in the literature. For example, \citet{ebcioglu1988expert} describes chord skeletons as sequences of rhythmless chords with fermatas, like the harmonic skeleton in~\citet{hild1992harmonet}. The actual notes including passing tones and suspensions are generated by a fill-in view object that takes the chord skeleton as input in~\citep{ebcioglu1988expert}.


\paragraph{Search space and context}

The complexity of the harmonization problem is defined by the size of the search space for viable chords. This size is in turn determined by the problem description, which includes the number of chord types and of chords to be generated. The size of the search space is relevant to the number of states in a hidden Markov models~\citep{allan2004harmonising}, the number of nodes in a neural network~\citep{hild1992harmonet}, and the length of the chromosome in a genetic algorithm~\citep{mcintyre1994bach}. 

In chorale harmonization, for a given key, the basic set of chords consists of: I, ii, iii, IV, V, V\textsuperscript{7}, vi, and vii\textsuperscript{o} and their positions (root or inversions). The size of the basic set is significantly increased if details such as secondary dominant, pivot chords, and key modulations are considered. The size of the search space can also be determined by examining composed examples.

Generating chorale harmonization can be approached as an iterative process. Given a melody, the system first generates possible configurations for the chord progression, then modifies the patterns based on certain criteria. For example, the expert system in~\citep{ebcioglu1988expert} takes the generate-and-test approach using three types of rules implemented as first-order logic: production rules, constraints, and heuristics. Systems that use genetic algorithms also follow this iterative nature: the ``chromosomes'' or ``population'' are modified iteratively to improve the quality based on fitness functions~\citep{mcintyre1994bach}. However, this iterative nature becomes computationally expensive when the chorales become longer. 

To overcome this problem of search space explosion, many researchers focus on local patterns instead of the entire compositions. This is not only a practical solution for computational reasons, but also a reasonable approach because many of the voice-leading rules in chorales are concerned only with local movements in and between individual voices. 

The modeling of local or short-term patterns is even more prominent in approaches that use neural networks and Markov models. In general, to determine a chord at the current time point, such systems define the local context by considering the recent chord sequences, and the melody note in the previous, current, and immediate future time points~\citep{allan2004harmonising}. For example,~\citet{de2010neural} discuss three models---one considering only the current chord, one incorporating the current and the immediately preceding chord, and one considering the current and the two closest preceding chords---and their combinations to determine the current chord. \citet{eigenfeldt2009realtime} also proposed a third-order Markov model for chord generation.

\paragraph{Cadences} 
While the cadence is a key harmonic feature used in the delineation of phrase boundaries, the modeling of cadences is not always explicitly addressed in chorale harmonization systems. Even for systems that account for phrase structure, cadences are handled to varying degrees of detail. 

The generation of cadences is typically achieved through constraints. For example, ending each phrase with a cadence can be set as a hard constraint for any chord progression~\citep{anders2009computational}. Cadences can also be induced through heuristics~\citep{ebcioglu1988expert} or preferences, say, in cost functions \citep{phon1999four, mcintyre1994bach} that bias the system towards producing more desirable cadential patterns. For example, in~\citep{phon1999four}, wrong cadences are penalized up to 100 points, 10 times more than any other rules governing voice leading, while \citet{mcintyre1994bach} awarded points for proper tritone resolution, including transitions from V\textsuperscript{7} and vii\textsuperscript{o} to I or vi.  

In the work of \citet{tsang1991harmonizing}, cadence formation is realized through four rules (out of a total of nine). Also using a rule-based approach, \citet{geis2007ant} included a resolution rule as a part of the harmonic score calculation. The generated cadence is then constrained through the rules to be similar to the chosen style. The modeling of cadences as constraints or preference rules can be readily incorporated in systems that use combinatorial approaches such as genetic algorithms and constraint programming. 

In contrast, cadential closure is discussed almost as a by-product in systems using statistical approaches such as neural networks and Markov models. For example, in~\citep{hild1992harmonet}, harmonic closure relies on explicit coding of the beginnings and endings of phrases. 
\citet{allan2004harmonising} report that their HMM with the Viterbi algorithm generates plausible cadences similar to those in the chosen corpus; little information is provided regarding how the system ensures correct cadences, especially the ones midstream, when seeking the most likely chord progression. 

Recently, \citet{Yi2007Automatic} proposed an interesting Markov model-based approach: instead of generating the most probable sequence, the authors modeled the harmonization problem as a Markov decision process so that sequences with the highest rewards, including those considering cadences, are selected. The reward is produced by a utility function, which can be either formulated based on music theory or learned from a dataset. In~\citep{Yi2007Automatic}, only two rules are encoded in the utility function: chords for which melodic notes are chord tones are preferred, and authentic cadences are preferred while plagal cadences are acceptable.

\subsubsection{General harmonization}
\label{sec:harm}

The general harmonization problem can be considered as one of determining multiple synchronized-note events to fit certain user-defined criteria. Compared to chorale harmonization, the problem of generating harmonic sequences in other genres is less well defined. Unlike chorales in which fitness functions can be established based on well-studied music theoretic rules, the style, cadences, harmonic quality, and even chord labels can often be unclear in the general harmonization problem.


A number of studies focus on generating harmonizations to user-created melodies in a popular style. Most studies adopt data-driven approaches to determine possible chords for a given melodic segment and to ensure chord-to-chord transitions are commonly observed in the examples. The systems typically produce a harmonic skeleton and use predefined patterns for creating rhythmic textures and instrumental arrangements. 


\citet{lee2004ring} used the first-order Markov model with dynamic programming to determine the harmonic skeleton for a user-hummed tune; the state transition probabilities are learned from 150 songs. ~\citet{simon2008mysong} took a similar approach; training their system on 298 songs from various genres such as jazz, rock, pop, blues, and others. Both systems are evaluated via subjective feedback from listening experiments. A drawback of this approach is that the chord sequences generated tend to be generic and indistinct in style.

To preserve a recognizable style, rather than training on multi-style datasets, \citet{chuan2011generating} focused on music composed by only one artist/band, or even a single piece in the extreme case; the problem of data sparsity is overcome through a chord tone determination module that generates a set of possible chords, and the use of neo-Riemannian operations to fill in missing transitions between chords. The generated chord sequence was evaluated subjectively, and quantitatively using cross entropy. 

A system for reharmonizing uplifting trance music based on a newly generated chord sequence was developed by~\citep{bigo2015viewpoint}. The system was extensively tested in an empirical study, in which~\citet{agres2016harmonic} found that repetitiveness in harmonic structure and tension, not solely rhythmic structure, is a contributor to listener enjoyment in this form of electronic dance music. 

\subsubsection{Jazz Chord Sequences}
\label{sec:chord_seq}

The problem of generating chord sequences unconstrained by melodic considerations is more frequently seen in jazz. Research on the generation of jazz chord progressions has focused on chord substitution and variation.

\citet{steedman1984generative} studied 12-bar blues and defined a small set of rules using a generative grammar that produces recognizable 12-bar blues chord progressions. As noted in the article, there was no explicit attempt to generate good chord progressions or to avoid bad ones. Instead, \citeauthor{steedman1984generative} examined the harmonically meaningful chord progressions and substitutions in order to generate sequences to accompany melodies.

\citet{chemillier2001improvising} provides a scenario in which a chord sequence of $n$ bars is repeated as a loop with variations as a foundational jazz accompaniment to explain why identifying substitutions to the original sequence is crucial in jazz improvisation. Although the substitution module in Chemellier's system randomly applies Steedman's rules to the chord sequence to generate variations, the author suggests ways that the user could interact with the system to steer the selection.

In the next section, we discuss systems for rhythm generation, without regard for pitch.

\subsection{Rhythm}
\label{sec:rhythm}

This section discusses systems that automatically generate rhythm. While some of the above mentioned systems already include aspects of rhythm, such as duration, the focus of this section lies on research that focuses on music generation for percussion instruments. 
In music generation systems, rhythm is often considered as given or embedded as an attribute of note events. Overall, there exists far fewer systems that solely generate rhythm than systems focusing on melody and harmony, but similar modeling approaches have been applied to rhythm generation. 

\citet{tokui2000music} proposed the CONGA system which combines genetic algorithms and genetic programming to produce rhythmic patterns that evolve, with user feedback as fitness function. Short fragments of rhythmic patterns form the chromosome elements in the genetic algorithm; the manner in which these patterns are concatenated into a sequence is determined by genetic programming. For evaluation, participants use the system to produce rhythmic progressions that sound like rock n' roll.

A genetic algorithm was also implemented by~\citet{ariza2002prokaryotic} to generate rhythms that consists of a sequence of genetic variations. His fitness function consists of calculating the distance between a rhythm and a user-provided ``fit-rhythm'' through five distance measures. The results were not evaluated in the paper. 

\citet{tidemann2008drum} used hidden Markov models to learn and generate core patterns and variations similar to examples played by different drummers. Core patterns and variations are defined by the supermaximal repeats (i.e., a repeated pattern that is not part of another pattern) in the melody that correspond to structural parts such as the verse, chorus, and bridge. To produce more ``human-like'' drum patterns, note onset times and velocities are modeled as Gaussian distributions with noise. The generated patterns were evaluated via a classification task to determine if the generated patterns belong to the same class as the training corpus.

\citet{hawryshkewich2011beatback} also applied statistical approaches to generate rhythmic patterns. The Beatback system uses variable-length Markov models to store users' input via a MIDI drum interface and to generate rhythmic patterns consistent with the users' styles and pattern complexity. Each drum event is described by its duration, velocity, and instrument such as hi-hat or snare; drum patterns are generated by reproducing highly-likely sequences as observed in the user's playing.


\citet{marchini2010unsupervised} used variable-length Markov models to generate percussion sequences, while their system learned and reproduced sequences from audio examples. Percussion sounds are first segmented into note events using onset detection; each event is then mapped to symbolic sequences via hierarchical clustering based on acoustic similarity. Preference is given to symbolic labels that maximize temporal regularity. To generate future events that respect the metrical structure, a temporal grid is created via beat and tempo detection. 

Similarity in a social context is explored by \citet{weinberg2006toward}, who created Haile, an interactive robot that plays the drums. Haile analyses, in real time, perceptual aspects of human players, and decides on one of six interaction modes in which to generate rhythms to play with the human player based on this analysis. The six interaction modes are: imitation, stochastic transformation, perceptual transformation, beat detection, simple accompaniment, and perceptual accompaniment.


The next section moves away from typical music generation systems and deals with the more complex tasks of orchestrating music and of accounting for timbre in music generation.

\subsection{Timbre}

This section considers the aspect of timbre in music generation. The timbre, also referred to as tone color of sound, is the property that distinguishes different voices and musical instruments. In the context of music generation it forms an important aspect to consider when, for instance, composing music for an orchestra because the timbre of each individual voice has an effect on the perception of the composite sound. 


The problem of orchestration with a target timbre is often modeled as a combinatorial problem in which the system aims to search a database of instrument sound samples to retrieve some combination of a subset of the sounds that produce a similar perceived timbre.

In early systems, timbral closeness was only measured using similarity in the frequency spectrum. For example, \citet{psenicka2003sporch} proposed the SPORCH system, which provides orchestration for any acoustic instrument ensemble to match as best possible any arbitrary sound file. The database consists of descriptive instrument features such as pitch range, the loudest/softest dynamic levels, and notation information (e.g. clef, transposition, etc.) about sound samples. Orchestration is then determined through iterative search for the best instrument sample mix with frequency spectral peaks most similar to those of the target file.

Similarly, in~\citet{hummel2005simulation}, the system searches iteratively for virtual music instruments that when synthesized together create speech-like timbres. The algorithm minimizes the difference in spectral envelope between the current sound and the target timbre by iteratively adding sounds that minimize the residual error.

\citet{mccormack1996grammar} developed an L-grammar that generates polyphonic music. Their model learns features based on pitch, duration and timbre. The details of the timbral characteristics were not disclosed in the paper. 

\citet{carpentier2010solving} points out that acoustic instrument orchestration is significantly different from, and more complex than, sound synthesis. They model orchestration as a constrained multi-objective search problem wherein the system aims to find combinations of sounds similar to a target timbre. Sound samples stored in a database are represented by their sound attributes and features. Sound attributes are symbolic labels related to discrete variables in compositions, including pitch, dynamics, and playing style; sound features represent psychoacoustic characteristics such as brightness and roughness that can be used to quantify perceptual dissimilarity. To minimize the perceived dissimilarity, the authors used randomly weighted Chebychev aggregation functions to model dissimilarity as a set of mono-objective problems. Finally, a genetic algorithm is employed to find the optimal combination given constraints on sound attributes as well as perceptual dissimilarity.

More recently, \citet{collins2012automatic} applied machine learning to automatic composition of electroacoustic music, taking into account the quality of the final mix. A piece is composed by combining and modifying existing audio segments. The system first analyses the features of the audio segments, such as percussive onset patterns and absolute peak amplitude, to produce suitable intermediate material for mixing. The segments are then modified by applying effects including delays, filters, and time stretching. The composition is iteratively refined using a density envelope on structural parameters such as perceived loudness, sensory dissonance, and increased tension to control the level of activities. The best mix is selected as the one most similar to an exemplar piece using dynamic time warping. A similar approach is employed by~\citet{sturm2006adaptive}, who uses adaptive concatenative sound synthesis to generate and transform digital sound. Short segments are used to synthesize variations of sound much like a collage, based on a measure of similarity, the L$_1$-norm of the difference, of audio features. 

In the next three sections, aspects newly associated with music generation are discussed, such as interaction, narrative and difficulty. The section to follow most immediately will explore interactive improvisation systems which are capable of performing together with a human player. Style replication in real-time jazz and other improvisation systems is also addressed in the upcoming section.

\subsection{Interaction}
\label{sec:dialog}

This section considers systems in which two-way communication between the computer and human player(s) exist; both the player and the system listen to what is being played, anticipate, and improvise new music in real time. 
Previous examples addressed music generation in the absence of live interaction with a player, based on the generated music's similarity to either a target style or piece. Here, we focus on similarity in a social context and turn to interactive systems, in which the generation algorithm ``improvises'' in real-time with a player. In this scenario, similarity to the style of the player and self-similarity to what is previously played within the piece become an important goal, thus shifting the focus to the requirements of interaction. 

While there are computer-assisted composition systems that allow the user/composer to interact with the system and iteratively improve generated solutions in a non-performance setting (e.g. \citep{farbood2007composing}, most of these systems have been discussed in their respective sections above. In this section the focus lies on real-time \emph{performance} systems.

\paragraph{Early Work}
One of the earliest automatic improvisation systems was created by George Lewis in the 1980s. One of his compositions, called Voyager, is composed in automatic response to a musician playing, as well as to the program's ``own internal processes''. In this early work, the performer is not able to control the system during performance~\citep{lewis2000too}.

\paragraph{Structured improvisation} One of the first interactive jazz solo generators, GenJam \citep{biles1998interactive}, generates melody lines over a given chord progression. It listens to a human player's last four bars, maps it to a chromosome representation and evolves what it ``hears'' with a GA into what it will play in real time. Fitness evaluation is performed by a human listener who continually gives feedback, rating the output as ``good'' or ``bad''.

\citet{thom2000bob} created Band-out-of-the-Box (BoB), an agent built for interactive jazz/blues improvisation of four-bar solos, with the goal of  developing a system that is realistic and fun to play with. 
A probabilistic approach is used, based on variable tree encoding with multiple features---pitch class, interval, and melodic direction. The model is trained on warm-up sequences prior to the performance; the features extracted in the warm-up are first clustered based on histograms; the resulting statistics are then used during real-time generation to determine the current musical environment. 

Interactive jazz generation is explored further in research by \citet{franklin2001multi}, who uses a set of rudimentary rules for jazz and a neural network in combination with reinforcement learning to trade fours between a musician and the system. The system, called CHIME, has a stochastic element that allows for out-of-chord changes, which the author suggests can be done more pointedly and purposefully in future research. The author also points out that the hard coded rules do not encompass the developing of a statement or the creation of a shape.


\paragraph{Free improvisation}

A second type of improvisation system generates music more freely with a performer, in real-time, without a fixed, predefined structure. 
\citet{pachet2003continuator}'s Continuator uses a Lempel-Ziv parsing algorithm~\citep{assayag1999guessing}---adapted to properly handle rhythm, beat, harmony and imprecision---to learn the characteristics of any style. It is able to concurrently learn and generate a stream of music that is similar to a style such as jazz or a player's own style; it can generate music, either as a standalone system, as continuations of a performer's input, or as an interactive improvisation backup. By aggregating clusters of notes and treating them as units, the Continuator is able to handle polyphonic music. 

Using a different data structure, the factor oracle, improvisation systems belonging to the OMax family~\citet{assayag2006omax} can also concurrently encode and generate music in a  player's style, and handle polyphonic music. The factor oracle is a finite state automaton, originally designed to  efficiently search for substrings (factors) in a text~\citep{allauzen1999factor}. More recent extensions further allow the OMax system to handle audio signals instead of symbolic MIDI; the resulting audio-based system is called Ofon. OMax's approach has also been applied to speech to simulate rap, and to video frames~\citep{bloch2008introducing}.

\citet{assayag2010interaction} and \citet{dubnov2012music} further experimented with audio-based factor oracles to improvise music resulting in systems such as the OMax-Ofon system developed by \citet{assayag2010interaction}. The system developed by~\citet{dubnov2012music} produces variations from an audio recording using a graph of repeated factors found in the recording; the system's main challenge consisting of the marking and allocating of regions in the original audio that are deemed most promising for the oracle to focus on in order to achieve the desired result. To do this, \citeauthor{dubnov2012music} introduce an analysis method based on Information Rate (a concept previously linked to musical anticipation~\citep{dubnov2006spectral}); in contrast to the previous factor oracle systems, the audio analysis is done in advance, and a performance is in a sense pre-planned. 

The earlier OMax systems are agnostic to rhythmic and longer-range structures.
\citet{nika2015guided}'s system ImproteK, also based on the factor oracle, generates both rhythms and harmonies. It was recently built into an architecture that aims at combining reactivity and anticipation in the music generation processes steered by a ``scenario'', which can be a four-bar jazz chord sequence. The system composes off-line given a scenario; during the performance, this can be re-written to fit the performance. 

\paragraph{Performer feedback}
Mimi~\citep{franccois2007visual}, like the OMax family of interactive improvisation systems, is based on the factor oracle. A novelty of this system is that it allows the user to visually see the recent past and future generated music, so as to be aware of the musical context and to plan a response. The performer is also the operator of Mimi, controlling her learning rate, switching on and off the learning, and clearing her memory~\citep{schankler2014improvising}. 
A variation on Mimi, Mimi4x~\citep{franccois2013mimi4x}, allows the user to control four interacting Mimi instances and to structure an improvisation by deciding when and which of the four Mimi-generated streams start and stop, their re-combination rates, and the playback dynamic levels.

\paragraph{Other interactions}
Besides the above-mentioned systems, in which music is generated in partnership with a human musician, some systems use other types of interactions. A system created by~\citet{marsden2004novagen} generates melodies based on the movements of a dancer, combined with elaborations based on Schenkerian analysis. The system is given a ``background'', which consists of one note per bar, key and meter information. Depending on the speed of the dancer's movements, more target notes are generated. Gait is another determining influence, for example, a crouching gait is associated with high regularity. The output of the system follows harmonic and intervallic patterns found in real music, yet lacks subdivision into meaningful phrases. The association, reflecting similarity, between the movements of a dancer and changes in melody is credible, although it does not convey the feeling that the dancer is controlling the music, mostly because of a slight lag in the system. The interactive music improvisation system (SICIB) developed by
\citet{morales2001sicib} has a similar setup, as it detect motion from sensors attached to dancers, and uses a rule-based approach to translate this to music. Motion characteristics such as curvature
and torsion of movements, and velocity and acceleration are taken into account. The generation is performed by the Escamol system which uses grammar-rules~\citep{morales1992non} and real-time synthesis by Aura~\cite{dannenberg1996flexible}

Another interactive system, the robot drummer called Haile, developed by~\citet{weinberg2006toward} is discussed in more detail in Section~\ref{sec:rhythm}, which calls out the rhythmic aspects of music generation systems. The next section will explore music with a narrative, which includes game music and video background music.

emotion


\subsection{Narrative}

Narrative music is music that tells a story. The narration consists of a set of representational, organizational, and discursive cues that deliver story information to the audience. In this section we focus on different types of narratives, such as tension profiles, the blending of fragments for game music, leitmotifs, and film music. 

Narrative cues create structure within music, including variation in emotions evoked throughout a piece (synchronized with video or game play), tension profiles, leitmotifs, repeated patterns and others. The enforcing of narrative structure can lead to similarity within a piece (e.g. repeated patterns and motifs) or, on a higher-level, between emotions evoked by the music and simultaneous media such as video or games.

In recent years, there has been increased interest in creating background music for games and video. The goal in these types of composition problems is that the music should match the content or emotional content of a scene/narrative. The idea of program music, music having an extra-musical narrative or purpose, is an old one. For example, ``the adventures of Don Quixote'' composed by Richard Strauss and Hector Berlioz's ``Symphonie fantastique'' both derive inspiration from extra-musical sources. 
Inherent in music with a narrative is the existence of long term structure, which was already touched upon in Section~\ref{sec:melody}; the discussion continues in the following paragraphs.

\paragraph{Tension}

An important tool for evoking emotion is the use of musical tension. Audio features, such as roughness, have been shown to correlate with perceived tension/relaxation patterns in music~\citep{vassilakis2005improvisation}. \citet{farbood2012parametric} conducted an extensive experiment to build a perceptual tension model that takes into account the dynamic, temporal  aspects  of  listening. Farbood models tension in terms of multiple musical parameters, inclusive of both audio and score-based features. \citet{farbood2007composing} also created Hyperscore, a graphical, computer-assisted composition system in which users can intuitively edit and visualize musical structures. Both low-level and high-level musical features (such as tone color, melodic shape, dynamics, harmonic tension) are mapped to graphical elements that users can control and which allows them to create compositions. This allows users to, for instance, draw a tension line for the new composition. 

\citet{browne2009global} used simulated annealing to arrange pre-written motifs according to a pre-specified musical tension profiles. The tension profiles were computed using an ANN model, and Kullback–Leibler divergence was employed to measure the distance between the desired and the observed tension profiles.

More recently,~\citet{herremans16orbel} used a tonal tension model based on the spiral array~\citep{herremans16tension} to calculate tension of a polyphonic piece. The algorithm, called MorpheuS, constrains the detected patterns and generates music that fits as best possible a given tension profile. This tension profile can be provided by the user or calculated based on a template piece. The generation process is guided by a variable neighborhood search algorithm. 

The breaking of rules that govern Western tonal music elicits tension. In~\citet{rutherford2002experiment}'s scary music study, more scary music is generated by breaking the Western tonal music rules. The results were verified by human listeners who noted the scariness dimensions of the generated music.


\paragraph{Blending}

Game music is most frequently generated by cross-fading between audio files each time the player shifts from one game state to another \citep{collins2008game}. An exception is the music for Depression Quest\footnote{\url{https://isaacschankler.bandcamp.com/album/depression-quest-ost}} which generates music dynamically as one moves through the different scenarios in the game. Brian Eno, a composer known for creating generative systems for ambient music, collaborated with Maxis/EA games to create a soundtrack generation system for the game ``Spore'', in which the music changes based on a gamer's style of play~\citep{johnson2006long}. Details of how these systems work are not publicly available. 
In traditional games that use cross-fading, however, it is not uncommon for the two fragments to clash rhythmically or harmonically. The clash can be ameliorated by techniques such as crossfading quickly, which can be distracting or jarring. 

\citet{muller2012data} devised an automatic DJ system for crossfading that ensures smooth blending, yet still requires the audio fragments to be harmonically and rhythmically similar. 
Smooth blending can be improved by restricting the range of allowed rhythms and harmonies; however, this would also restrict the musical variations and expressive capacity of the music. In order to solve this problem, \citet{prechtl2014algorithmic} created a real-time system that generates the music from scratch instead of using existing fragments. The music generation process uses a stochastic model and takes into account emotion parameters such as alarm or danger.

\paragraph{Leitmotifs}
The system developed by~\citet{brown2012mezzo} focuses on ``Leitmotifs'', short and distinctive musical fragments associated with game characters and elements, a strategy commonly employed in Western opera. Each of these motifs are embedded in different musical forms; each musical form is associated with different degrees of harmonic tension and formal regularity, thus conveying different amounts of ``markedness''. In combination, the leitmotifs and forms correspond to different states of the story of a game. See \citet{collins2009introduction} for a more complete overview of procedural music in games.



\paragraph{Film music}
Music with a narrative is frequently used as background music to films. The effect that music has on perceived emotion in film has been studied by~\citet{parke2007quantitative}. When mapping perceived emotion to a three-dimensional space of stress, activity, and dominance, the geometrical center of mass of the three perceived emotions (in this space) when experiencing film and music combined is found to be in between that of the participants who listened to music alone and watched film alone. In the study, the film clips were selected for their ambiguous meanings. \citet{prechtl2014methodological} argue for the need for thorough empirical evaluation when generating music purported to communicate particular emotions. 

\citet{nakamura1994automatic} created a prototype system that automatically generates sound effects and background music for short video animations. Music---harmony, melody and rhythm---is generated for each scene, taking into consideration the mood, the intensity of the mood, and the musical key used in the preceding scene. The characteristics and intensity of the movements on screen determine the sound effects. 
Another example application for video background music is MAgentA, created by~\citet{casella2001magenta}, the goal for which is to generate ``film-like'' music using the mood of the environment within which it is embedded. 

The final functional aspect in music generation systems, that of the instrument playing difficulty, is discussed in the next section. 

\subsection{Difficulty}

The difficulty of a piece of music refers to the level of skill required for a musician to play the piece. 
When automatically composing musical pieces, the manipulation of features such as melody, harmony, rhythm, and timbre often rise to the fore, and ergonomic goals such as ease of playing are often ignored. One could argue that if a model is trained on existing pieces using the appropriate feature set, a new piece that is sampled from this model should be equally playable. It would be interesting to explicitly measure difficulty to verify this causality. Thus, generating a piece of similar playability to pieces in a corpus, or of a predefined difficulty level for an instrument could be the main goal of a music generation system.

\paragraph{Music generation according to difficulty level}

\citet{tuohy2005genetic} developed a genetic algorithm that generates playable guitar music by minimizing hand and finger movements. More recently, \citet{mcvicarautoguitartab} automatically generated lead and rhythm guitar music in tablature notation, based on a given chord and key sequence.

\citet{sebastien2012score} implemented a system that measures the difficulty of a piano piece based on seven different characteristics including harmony, fingering, polyphony, and irregularity of rhythm. Such a system could easily be improved with systems for automatically computing piano fingerings~\citep{lin2006intelligent, de2012differential, balliauw15} and string instrument fingerings~\citep{sayegh1989fingering,radi2004path, radicioni2004algorithm}. 
The combination of fingering and difficulty evaluation systems with music generation systems provides an opportunity to evaluate pieces in a non-traditional, yet essential way. 


\section{Future challenges}

Over the last few decades, research in music generation has achieved tremendous progress in generating well-defined aspects of music such as melody, harmony, rhythm, and timbre. State-of-the-art statistical models, advanced optimization techniques, larger digital databases on which to train models, and increase in computing power have all led to the field producing better systems. Why then are we not using music generation systems in our day to day lives? The above survey shows that an important overarching challenge remains: that of creating music with long-term structure. 

Long-term structure, which often takes the form of recurring themes, motivs, and patterns, is an essential part of any music listening experience~\citep{lerdahl1983overview}. Recent music generation systems have tackled this challenge by constraining certain types of long-term structure, such as recurrent patterns~\cite{herremans16orbel}, form~\cite{tanaka2016babbitt,herremans2015generating}, cadence~\cite{cunha2016}, and pitch contour~\citep{pachet2001finite}. Secondly, developments in the field of deep learning~\citep{eck2002first,ICML2012BoulangerLewandowski_590} show that neural networks can incorporate memory structures when learning sequential data. The ability of techniques such as RNN and LSTM to capture long-term structure should be further investigated. 

In order to make computer generated music systems part of our daily lives, there is a crucial need for more ``intelligent'' systems in which newly composed music matches higher-level concepts. This intelligence can be expressed in the functional domain of the ``narrative''. While there are recent attempts at generating music with tension~\cite{farbood2007composing,herremans16orbel}, that matches a computer game~\cite{prechtl2014algorithmic}, that embody leitmotivs~\cite{brown2012mezzo} or that accompany film~\cite{nakamura1994automatic}, and others, there is still ample room for better understanding the connection between music and emotion, so as to integrate this crucial relationship in music generation systems. This could lead to real-life practical applications such as real-time music generation for games, and background music for film and video. 

While machine learning techniques can be extremely useful in tasks such as the above-mentioned modeling of emotion in music, they usually require large amounts of data. Therefore, the field has seen an ongoing need for more data. There lies a real potential for future work to move towards intelligent systems that do not require copious amounts of data, that are capable of innate reasoning, and thus better mirror the workings of the human mind. This would also solve the continuous challenge of finding a balance between regenerating existing music and novel fragments without plagiarizing, as touched upon in Section~\ref{sec:success}. 

One of the characteristics of computer generated music that is often neglected is playing difficulty. While one application would be to tailor novel music to a certain level of musician skill, there is also potential for using detected/calculated playing difficulty as an evaluation measure for generated music. 

A challenge related to making music generation systems usable by the general public is, not only the quality of the generated musical content, but also the quality of the rendering. While this is not an aspect that we explicitly surveyed in this paper, it nevertheless is important in creating a real-life applications. In recent years, the field of automatic music production has gained increasing traction. Research topics in this field include human-like rendering of midi files with expressive timing~\cite{bresin2000emotional, grachten2014analysis} and automatic mixing~\cite{deruty2016goal}. Furthering the development of systems for realistic rendering of generated music, which is often in MIDI, will stimulate the attractiveness and usability of music that is generated automatically. 

The success of music generation systems is not only measured through the practical adoption of the systems. Over the course of the years, researchers have adopted multiple methods for evaluating the output of systems, as outlined in Section~\ref{sec:success}. It remains difficult, however, to objectively compare different systems as they usually take different input parameters, generate different aspects of music, are trained on different styles, or do not have audio examples available for the reader. Furthermore, in listening experiments, the Mere-exposure effect~\citep{zajonc1968attitudinal} will make listeners prefer existing pieces over new ones, as familiarity causes a higher enjoyment. To address this need for the proper comparison of systems to assess the state-of-the art, the authors of this paper have set up a publicly accessible repository of computer generated music systems\footnote{\label{note:cogemur}http://dorienherremans.com/cogemur}. Apart from the goal of stimulating the visibility of music generation systems and their outputs, this online repository will facilitate the comparison of systems by collecting detailed information such as the nature of the system's input/output, and potential manual corrections performed.




\section{Conclusions}







This article has presented a taxonomy for the key concepts that form the functional goals of music generation systems. We then provided a survey of the state-of-the-art in music generation systems with respect to this functional taxonomy. 
By focusing on what current systems can and cannot do, rather than the algorithmic techniques, we obtain a clearer view of the frontiers of automatic music generation, thus setting the stage for new breakthroughs. This approach has allowed us to identify uncharted areas and challenges for the field of automatic music composition. 



In line with the current trend of companies such as Google (through the Magenta project) and Jukedeck, music generation systems will become ever more prominent in our day to day lives. The functional overview of systems described in this paper shows the areas with opportunities for further advancement to make automatic music generation a viable tool for applications ranging from artistic innovation to the creation of adaptive, copyright-free music for games and videos. Current challenges of the field include generating music with long-term structure; capturing higher-level content such as emotion and tension; creating models that possesses innate reasoning so as to reduce the amount of training data needed; and the promotion of transparent and objective evaluation methods. In order to facilitate the latter and stimulate visibility and evaluation of current music generation systems, the authors have set up an online repository for computer generated music results\textsuperscript{\ref{note:cogemur}}. 

\bibliographystyle{ACM-Reference-Format-Journals}
\bibliography{paper}

\received{August 2016}{000}{000}

\end{document}